\documentclass[11pt]{article}
\usepackage{graphicx}

\def\singleandabitspaced{\baselineskip=\normalbaselineskip\multiply
    \baselineskip by 110\divide\baselineskip by 100}
\def\singlespaced{\baselineskip=\normalbaselineskip}

\newcommand{\centeron}[2]{{\setbox0=\hbox{#1}\setbox1=\hbox{#2}\ifdim
                             \wd1>\wd0\kern.5\wd1\kern-.5\wd0\fi \copy0
                             \kern-.5\wd0\kern-.5\wd1\copy1\ifdim\wd0>\wd1
                             \kern.5\wd0\kern-.5\wd1\fi}}
\newcommand{\ltap}{\>\centeron{\raise.35ex\hbox{$<$}}
                     {\lower.65ex\hbox{$\sim$}}\>}
\newcommand{\gtap}{\>\centeron{\raise.35ex\hbox{$>$}}
                     {\lower.65ex\hbox{$\sim$}}\>}

\begin{document}

\singlespaced

\begin{titlepage}

\begin{center}
\vspace*{0.8in} \mbox{\Large \textbf{The Impact of Heavy Nuclei on the Cosmogenic Neutrino Flux}} \\
\vspace*{1.6cm} {\large Dan Hooper$^1$, Andrew Taylor$^{1}$ and Subir Sarkar$^2$ } \\
\vspace*{0.5cm}
{\it $^1$Astrophysics, University of Oxford, Oxford OX1 3RH, UK\\
     $^2$Theoretical Physics, University of Oxford, Oxford OX1 3NP, UK} \\
\vspace*{0.6cm} {\tt hooper@astro.ox.ac.uk, amt@astro.ox.ac.uk, sarkar@thphys.ox.ac.uk} \\
\vspace*{0.2cm}
Keywords: Ultrahigh energy cosmic rays, ultrahigh energy neutrinos \\
\vspace*{1.5cm}
\end{center}

\begin{abstract} 
\singleandabitspaced
 
As ultra-high energy cosmic ray protons propagate through the
universe, they undergo photo-meson interactions with the cosmic
microwave background, generating the `cosmogenic' neutrino flux. If,
however, a substantial fraction of the cosmic ray primaries are heavy
nuclei rather than protons, they would preferentially lose energy
through photo-disintegration so the corresponding neutrino flux may be
substantially depleted. We investigate this issue using a Monte Carlo
simulation of cosmic ray propagation through interagalactic radiation
fields and assess the impact of the altered neutrino fluxes on
next-generation neutrino telescopes.

\end{abstract}

\end{titlepage}

\newpage
\setcounter{page}{2}
\singleandabitspaced

\section{Introduction}

The origin of the highest energy cosmic rays is among the most
interesting puzzles of modern astrophysics and may hold clues to new
fundamental physics \cite{Nagano:ve,Anchordoqui:2002hs}. Both air
shower and atmospheric fluorescence experiments have detected
ultra-high energy cosmic rays (UHECRs) with energies up to and beyond
$10^{20}$ eV
\cite{Lawrence:1991cc,Bird:1994uy,Takeda:1998ps,Abu-Zayyad:2002sf}. If
these are protons, then their energies are well above the predicted
`GZK cutoff' \cite{Greisen:1966jv,Zatsepin:1966jv}. Additionally,
their sky distribution is isotropic and their arrival directions do
not correlate with any plausible nearby sources. This has prompted
many speculative models involving new physics, e.g. decaying
superheavy dark matter in the Galactic halo
\cite{Berezinsky:1997hy,Birkel:1998nx}. Alternatively, the UHECRs may
be produced in the local interactions of particles such as neutrinos
which can travel cosmological distances without interacting with the
cosmic microwave background (CMB) --- the `$Z$-burst' mechanism
\cite{Weiler:1997sh,Fargion:1997ft}. Even more exotic possibilities
have been considered, for example the violation of Lorentz invariance
at very high energies \cite{Kirzhnitz:1971,Sato:1972}.

Astrophysical solutions to this problem may also be viable. A
relatively local source could, in principle, be responsible for the
highest energy events observed (although no plausible sources have
been identified \cite{Anchordoqui:2002hs}) and the isotropic
distribution may be due to larger than expected intergalactic magnetic
fields. Alternatively, a substantial quantity of heavy nuclei (rather
than only protons) may be accelerated in the cosmic ray sources. Heavy
nuclei, with their higher electric charge hence smaller rigidity,
would be more strongly deflected by magnetic fields and thus would be
more likely to appear as an isotropic distribution of
events. Additionally, heavy nuclei propagate over cosmological
distances differently than protons, raising the possibility that they
could originate from more distant sources
\cite{Stecker:1998ib}. Moreover, due to their higher electric charge,
the `Hillas criterion' for the acceleration of heavy nuclei is
relaxed relative to protons \cite{Nagano:ve}.

Experiments studying UHECRs have limited ability to measure their
composition through comparison of the characteristics of the air
showers with Monte Carlo simulations based on empirical hadronic
interaction models \cite{Anchordoqui:2004xb}. Data from Fly's Eye
suggested a gradual transition from heavy primaries at
$\sim3\times10^{17}$ eV to proton domination at $\sim10^{19}$ eV
\cite{Bird:1993yi}. Its successor, HiRes, reported a similar result
\cite{Abu-Zayyad:2000ay}. Although this is consistent with the
indication from Haverah Park data that less than 30\% of cosmic rays
above $\sim10^{19}$ eV are iron nuclei \cite{Ave:2000nd}, in the
energy range $\sim2\times10^{17}-10^{18}$ eV, iron nuclei are found to
dominate 2 to 1 \cite{Ave:2002gc}. This is also indicated by a recent
reanalysis of Volcano Ranch data in the energy range
$\sim5\times10^{17}-10^{19}$ eV \cite{Dova:2003an}. Moreover, it has
been argued that the highest energy Fly's Eye event at
$\sim3\times10^{20}$ eV could have been a heavy nucleus
\cite{Halzen:1994gy,Anchordoqui:2000sm}.

It has been noted that the interactions of UHE protons propagating
over cosmological distances should generate a flux of neutrinos
\cite{Berezinsky:1970xj,Stecker:1978ah,Hill:1983xs} and much effort
has been devoted to calculating this `cosmogenic' neutrino flux
\cite{Engel:2001hd,Fodor:2003ph}. This is often thought of as a
``guaranteed'' source of UHE neutrinos, expected to be detectable in
the next generation of experiments such as IceCube
\cite{Ahrens:2003ix}, Auger \cite{augertau} and Anita \cite{anita}. It
is clearly important to determine how much this flux would be altered
if the cosmic ray spectrum contains a substantial fraction of heavy
nuclei.

\section{Propagation of Ultra-High Energy Protons and Heavy Nuclei}

To study UHE heavy nuclei propagation, we have constructed a Monte
Carlo program to model the relevant interactions. Our simulation
includes the continuous energy losses associated with pair production
and the adiabatic expansion of the universe, as well as the key
stochastic processes --- photo-meson interactions of nucleons and
photo-disintegration of heavy nuclei.

At very high energies, protons interact with the CMB producing
electron-positron pairs, $p + \gamma_{\rm{CMB}} \rightarrow p + e^+ +
e^-$. The energy loss rate for this process climbs rapidly until a few
times $10^{19}$ eV where it reaches its maximum. The pair production
energy loss length in the range $10^{19}-10^{20}$ eV is a few hundred
Mpc \cite{pair}. This process can safely be treated as continuous for
our purposes.

The situation is somewhat different for heavy nuclei. The energy loss
rate due to pair production is proportional to the charge of the
nucleus squared, and thus can be considerably more rapid. The
threshold energy for this process is also higher, however, e.g. the
energy loss rate for iron nuclei peaks above $10^{21}$ eV, much higher
than for protons \cite{pair}.

More important for our purposes than energy losses due to pair
production are the effects of photo-meson interactions and
photo-disintegration. At these energies, processes such as $p +
\gamma_{\rm{CMB}} \rightarrow p + \pi^0$ and $p + \gamma_{\rm{CMB}}
\rightarrow n + \pi^+$ can take place via the exchange of a
$\Delta$-hadron near resonance (1.232 GeV center-of-mass energy). The
cross-sections for these processes are very large, leading to energy
loss lengths of tens of Mpc for protons above a few times $10^{19}$
eV. It is this effect that is expected to produce the GZK cutoff in
the cosmic ray spectrum \cite{Greisen:1966jv,Zatsepin:1966jv}.

Whereas the most important energy loss process for protons is the
photo-production of pions, for ultra-high energy heavy nuclei, it is
the process of photo-disintegration which has a much lower threshold
\cite{disintegration,Stecker:1998ib}. Here the nucleus interacts with
a background photon causing it to break up into a lighter nucleus or
nuclei along with a (typically) small number of protons and/or
neutrons. In part of the energy range we are interested in, the
dominant opacity for this comes from scattering of heavy nuclei on the
cosmic infra-red background (CIB) rather than on the CMB. However, in
contrast to the latter, the spectrum of the CIB is not well measured,
leading to considerable uncertainties in the propagation of heavy
nuclei. The cross-sections for N-nucleon emission for a given nucleus
have been experimentally measured in the energy range we are
interested in and we use the parameterizations given in
Ref.~\cite{heavysigma}.

\section{The Generation of Cosmogenic Neutrinos}

When the UHE primaries are protons, the main source of cosmogenic
neutrinos are charged pions produced in photo-meson
interactions. These decay as $\pi^+ \rightarrow \mu^+ \nu_{\mu}
\rightarrow e^+ \nu_e \bar{\nu_{\mu}} \nu_{\mu}$, providing a rich
source of neutrinos at EeV energies. Neutrons are created through $p +
\gamma_{\rm{CMB}} \rightarrow n + \pi^+$ and their decays generate
more neutrinos, $n \rightarrow p^+ + e^- + \bar{\nu_e}$. The energy of
these neutrinos is considerably smaller, however, producing a
secondary flux peaking in the PeV range.

\begin{figure}[t]
\centering\leavevmode
\mbox{
\includegraphics[width=4.2in,angle=90]{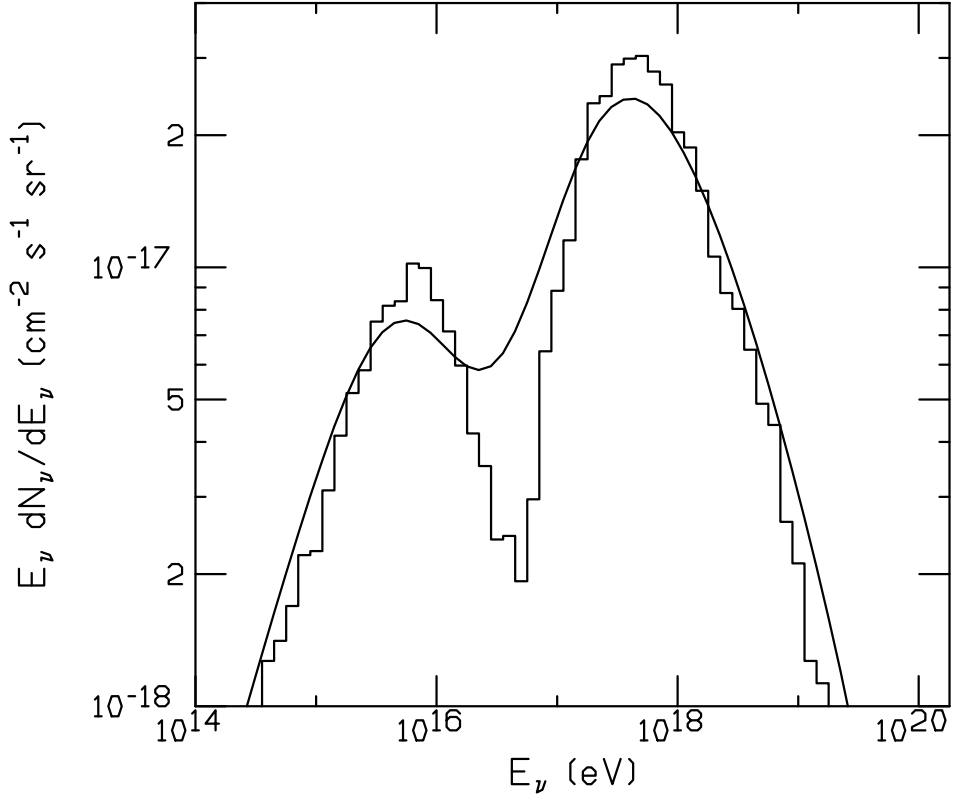}
}
\caption{The cosmogenic neutrino (plus anti-neutrino) flux produced in
the propagation of ultra-high energy protons (histogram). The solid
curve shows a previous result \cite{Engel:2001hd} for comparison.}
\label{cosflux}
\end{figure}

To calculate the cosmogenic neutrino spectrum, the injection spectrum
of protons (or heavy nuclei) and the cosmological distribution of
sources need to be specified. To enable comparison with previous work
\cite{protonprop}, we adopt the popular parameterizations
\cite{Waxman:1998yy}:
\begin{equation}
\frac{{\rm d}N_p}{{\rm d}E_p} \propto E_p^{-2}\,  
 \exp\left[\frac{-E_p}{10^{21.5}\, \rm{eV}}\right],
\label{spec}
\end{equation}
and
\begin{eqnarray}
H (z) &=& (1+z)^3 \hspace{3cm} \rm{for} \,\, z < 1.9, \nonumber \\
      &=& (1+1.9)^3 \hspace{2.7cm} \rm{for} \,\, 1.9 < z < 2.7, \nonumber \\
      &=& (1+1.9)^3 \times {\rm e}^{(2.7-z)/2.7} \hspace{5mm} \rm{for} \,\, z > 2.7, 
\label{dist}
\end{eqnarray}
for the cosmic ray injection spectrum and source distribution,
respectively. With these choices, we use our Monte Carlo to calculate
the cosmogenic neutrino flux for proton primaries shown in
Figure~\ref{cosflux}. We find reasonable agreement with the previous
calculation made using the same injection spectrum and source
distribution, but with a more sophisticated treatment of photo-meson
interactions \cite{Engel:2001hd}. For further details concerning the
interactions UHE protons undergo during propagation, we direct the
reader to Ref.~\cite{protonprop}.

The neutrino flux produced in the propagation of heavy nuclei can be
quite different from the result shown in Figure~\ref{cosflux}. As
nucleons are broken off from a heavy nucleus, the Lorentz factor is
unchanged so their energy is smaller by a factor of the nuclear mass
number, $A$, than the energy of the primary cosmic ray. Therefore,
considerably more energetic heavy nuclei are required to produce
neutrinos via charged pion decay. For iron nuclei ($A=56$), energies
of $\sim 2 \times 10^{21}$ eV or higher are needed to produce
neutrinos by this mechanism. Given the much smaller flux injected at
such high energies, the neutrino flux from charged pion production can
be considerably depleted if heavy nuclei make up a substantial
fraction of the ultra-high energy cosmic rays. It is easy to see that
for a differential $E^{-2}$ spectrum, the result in
Figure~\ref{cosflux} represents an {\em upper bound} on the EeV
neutrino flux from pion decays if the primaries are heavy nuclei.

The same is not true for neutrinos produced in neutron decays. As
heavy nuclei propagate and undergo photo-disintegration, both protons
and neutrons are removed and the neutrons decay producing neutrinos as
described earlier. Thus for each heavy nucleus, it is possible to
generate up to 3 neutrinos per neutron ($A-Z$), e.g. up to 90
neutrinos for each iron nucleus. In practice, many nuclei do not fully
disintegrate, however, and thus generate fewer neutrinos than might be
expected through this process.

The rate at which heavy nuclei photo-disintegrate depends strongly on
the density of CIB photons. Given the large observational
uncertainties in the CIB fluxes, this can lead to ambiguities
regarding the propagation of heavy nuclei. In particular, it is
difficult to predict the shape of the propagated cosmic ray spectrum
at Earth for heavy nuclei primaries with much confidence. This
introduces some uncertainty in the normalization of the primary
spectrum.

As described earlier, the neutrino spectrum in Figure~\ref{cosflux} is
produced by two mechanisms --- charged pion and neutron decay,
corresponding to the higher and lower energy peaks, respectively. For
UHE heavy nuclei to produce neutrinos in the higher energy peak, the
protons created by their photo-disintegration must have energies above
the GZK cutoff. This requires a nucleus with an energy of roughly $A
\times 4 \times 10^{19}$ eV (alternatively, a Lorentz factor of a few
times $10^{10}$). For such energetic nuclei, interactions with the
CMB, and not the CIB, dominate \cite{disintegration}. Only at lower
energies, where interactions produce sub-GZK protons, does the CIB
provide the main target.

Thus the neutrino flux populating the higher energy peak does not
depend directly on the assumed CIB spectrum although there is an
indirect effect because the propagated cosmic ray spectrum has to be
correctly normalized. The lower energy peak does, however, depend
directly on the intensity of the CIB. Heavy nuclei with energy around
$A \times 10^{19}$ eV or lower begin to photo-disintegrate due to
interactions with CIB photons, producing neutrons which decay
generating neutrinos in the lower energy peak. However this lower
energy population of neutrinos is considerably more difficult to
detect experimentally due to the much larger backgrounds. For this
reason, we will focus our attention primarily on the neutrinos
generated in the decay of charged pions and will not be overly
concerned with the debates regarding the intensity of the CIB. We
adopt the spectrum proposed in Ref.\cite{Malkan:2000gu}, which is
broadly consistent with TeV $\gamma$-ray observations of active
galactic nuclei. If we use instead the compilation of direct
measurements of the CIB \cite{Aharonian:2003wu}, the neutrino fluxes
in both peaks increase by a factor of $\sim 2$. Given that some of
these CIB measurements may be contaminated and thus represent upper
bounds to the cosmic flux, we do not pursue this issue further.

\begin{figure}[t]
\centering\leavevmode \mbox{
\includegraphics[width=4.2in,angle=90]{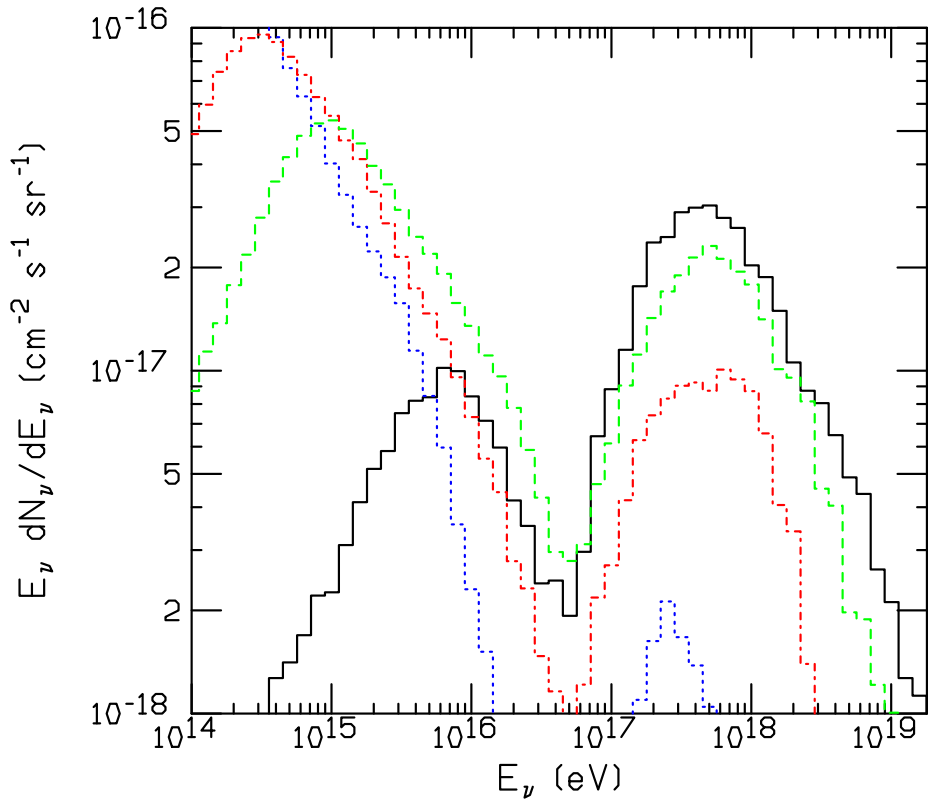} }
\caption{The neutrino spectrum produced in the propagation of the UHE
 heavy nuclei --- $^4$He (green, dashed), $^{16}$O (red, dot-dashed)
 and $^{56}$Fe (blue, dots) --- compared to the result for protons
 (black, solid line).}
\label{neuflux}
\end{figure}

\section{Results}

The cosmogenic neutrino fluxes for iron, oxygen and helium primaries
are shown in Figure~\ref{neuflux}, along with the result for protons
for comparison. The curves have been normalized by matching the proton
plus heavy nuclei flux to the observed cosmic ray spectrum at $3 \times
10^{19}$ eV \cite{Nagano:ve}; this ensures that at higher energies,
the energy spectrum at Earth is bounded between the results announced
by AGASA \cite{Takeda:1998ps} and by HiRes
\cite{Abu-Zayyad:2002sf}. As expected, the higher energy population of
neutrinos (produced in charged pion decay) is depleted for heavy
nuclei, while the lower energy population of neutrinos (produced in
neutron decay) is enhanced. Unfortunately, the lower energy population
is considerably more difficult to detect experimentally (see \S 5).

\begin{figure}[t]
\centering\leavevmode \mbox{
\includegraphics[width=4.2in,angle=90]{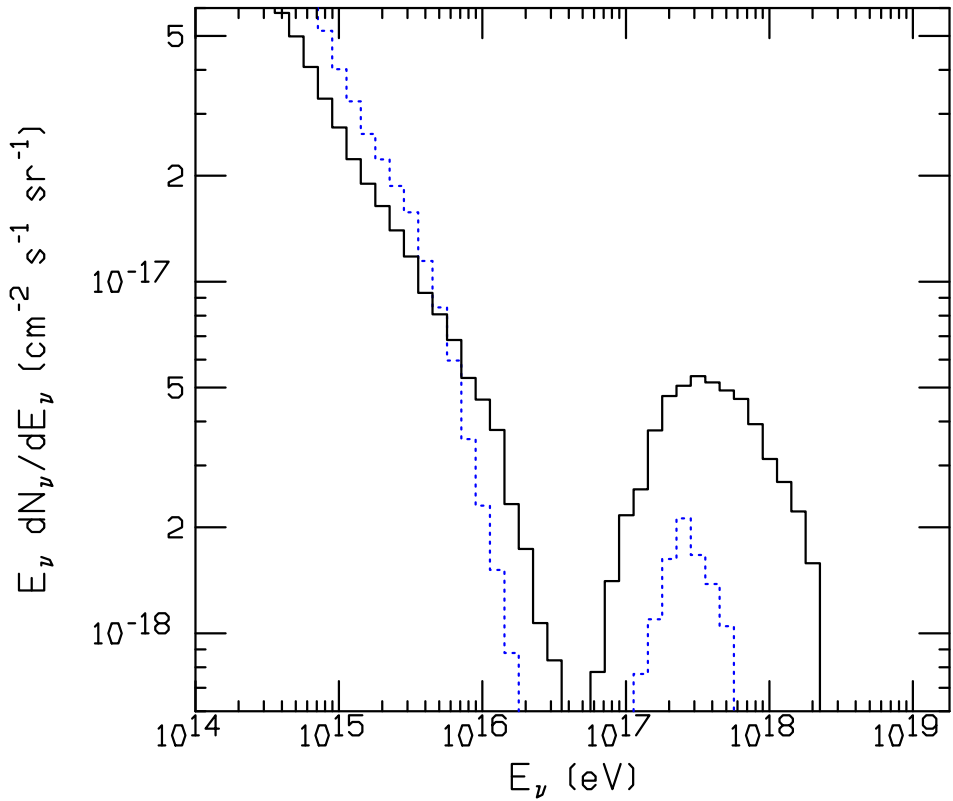} }
\caption{The neutrino spectrum produced in the propagation of
$^{56}$Fe nuclei. The dotted blue histogram results from the injection
spectrum of Eq.~(\ref{spec}), which has an exponential cutoff at
$10^{21.5}$ eV while the solid black histogram has the cutoff extended
to $10^{22.5}$ eV.}
\label{cutoff}
\end{figure}

These results depend on the assumptions made. In particular, the
neutrino spectrum can be considerably different if the observed UHECRs
come from a small number of powerful, nearby sources, rather than the
cosmologically distant distribution of Eq.~(\ref{dist}). The choice of
injection spectrum also plays an important role. We have assumed the
flattest possible power-law spectrum (\ref{spec}) with an exponential
cutoff imposed at $10^{21.5}$ eV. If the spectrum were to extend to
higher energies, more heavy nuclei would photo-disintegrate to yield
protons above the GZK cutoff and, hence more EeV-scale neutrinos would
be produced. In Figure~\ref{cutoff} we show that the flux is boosted
by about a factor of 3 if the cutoff is shifted upwards by a factor of
10.

\section{Event Rates in Next Generation Neutrino Experiments}

UHE neutrinos of cosmic origin have not yet been experimentally
detected but there are general expectations for significant fluxes
from various astrophysical and cosmological sources \cite{review}.
Several innovative experiments are currently being developed with the
sensitivity required to observe cosmogenic neutrinos within the next
few years --- these include IceCube, Auger and Anita.

IceCube, presently under construction at the South Pole, will be the
first km-scale neutrino telescope. It will be capable of observing
both muon tracks produced in charged current muon neutrino
interactions and shower events produced in charged and neutral current
events. Although optimized for the energy range of $10^{11}-10^{18}$
eV \cite{Ahrens:2003ix}, IceCube will be sensitive to neutrino
energies up to $\sim 10^{20}$ eV \cite{Alvarez-Muniz:2000es}.

The Pierre Auger Observatory, designed primarily to study the UHE
cosmic ray spectrum, will also be sensitive to UHE neutrinos. With
their much smaller interaction cross-sections, neutrinos would
interact well within the atmosphere and thus be identified as
`deeply penetrating, quasi-horizontal showers'. Earth-skimming tau
neutrinos would also provide distinctive detection signatures in Auger
\cite{augertau}.

The Anita experiment will be flown on balloons around the South Pole
using radio antennae to detect the interactions of UHE Earth-skimming
neutrinos in the polar icecap \cite{anita}.

Despite the very different techniques used by these experiments, all
three expect similar sensitivities to UHE neutrinos, in neighbouring
energy bands. About 1 UHE neutrino event per year is expected in
either IceCube or Auger given the cosmogenic neutrino flux of
Ref.~\cite{Engel:2001hd}, with a similar number expected in a 10-day
Anita flight. Given this, it is unrealistic to expect any of these
experiments to measure the cosmogenic neutrino spectrum in detail. Due
to the very low background at EeV energies, however, only a small
number of events is required to achieve a definite detection.

For concreteness, we have focused on the IceCube experiment and the
expected event rates are shown in Table~1. It is clear from these
numbers that substantial heavy nuclei content in the UHECR spectrum
can sharply reduce the ability of neutrino telescopes to observe the
cosmogenic neutrino flux. To improve on the sensitivity of present
experiments, new detection techniques may be required. This might
include extensions of IceCube \cite{icecubeplus}, space based air
shower experiments such as EUSO \cite{euso}, acoustic detection
technology \cite{acoustic}, or an experiment such as SALSA
\cite{salsa} using salt domes as a natural Cerenkov medium.

Notice in Table 1 that lowering the muon threshold of a detector from
1 PeV to 10 TeV results in only a fairly mild change in the predicted
event rate. Both the neutrino-nucleon cross-section and the energy
loss distance of muons in ice or water decrease with energy, making
neutrino telescopes considerably more efficient at higher
energies. Furthermore, the atmospheric neutrino background rapidly
falls off in this energy range. After the consideration of these
features, we conclude that, even in the case of iron nuclei, the lower
energy peak of Figures 2 and 3 will not be observable in any planned
experiment.

\vspace{0.0cm}
 \begin{table}
 \label{table}
 
 \hspace{1.5cm}
 \begin{tabular} {c c c c c} 
 \hline \hline
 & Showers & Muons ($E_{\mu}^{\rm{thr}}=1$ PeV) & Muons ($E_{\mu}^{\rm{thr}}=10$ TeV) & \\
 \hline \hline
Protons ($A=1$)  & 0.57 & 0.72 & 1.16 &\\
Helium ($A=4$)   & 0.42 & 0.50 & 0.80 & \\
Oxygen ($A=16$)  & 0.19 & 0.23 & 0.73 &\\
Iron ($A=56$)    & 0.036 & 0.042 & 0.17 &\\
 \hline \hline \end{tabular} \caption{The expected cosmogenic
 neutrino-induced event rates per year in IceCube, from the
 propagation of UHE protons and heavy nuclei which generate the
 neutrino fluxes shown in Figure~\ref{neuflux}. Rates are shown for
 both shower and muon events. For the former, a 1 PeV shower energy
 threshold was imposed, while for the latter muon energy thresholds of
 both 10 TeV and 1 PeV were considered.}  \end{table}

\section{Conclusions}

If the UHE cosmic rays are protons, the spectrum of neutrinos produced
during their propagation over cosmological distances is expected to be
detectable in next-generation experiments. However if the highest
energy cosmic rays are heavy nuclei, this neutrino flux may be
substantially depleted. For relatively light nuclei, e.g. $^4$He, the
suppression is only by about 50\%, but for heavier nuclei such
as $^{16}$O or $^{56}$Fe, the cosmogenic neutrino flux is reduced by a
factor of between 3 and 15. The reduction would be less severe if the UHE
cosmic ray spectrum extends beyond ZeV energies without
attenuation. Also of course the reduction would be at most by a factor
of 3, if a third of the UHE cosmic ray primaries are in fact protons.

Such a reduction in the cosmogenic neutrino flux would naturally make
experimental detection more difficult. On the other hand, a detection
of EeV neutrinos in experiments such as IceCube, Auger and Anita would
provide a complementary probe of the composition of UHE cosmic
rays. Alternatively if these experiments detect much larger UHE
neutrino fluxes than expected from the intergalactic propagation of
cosmic rays, then this would implicate local sources of UHECRs such as
decaying dark matter in the Galactic halo
\cite{Birkel:1998nx,Gondolo:1991rn,Barbot:2002kh}.

\section*{Acknowledgments}

As we were completing this work, we learnt of a similar study being
conducted by M.~Ave, N.~Busca, A.~Olinto, A.~Watson and
T.~Yamamoto. We wish to thank Igor Liubarsky and Dave Waters for
discussions. DH is supported by the Leverhulme Trust and AT holds a
PPARC studentship. We acknowledge travel support from the European
Science Foundation's ``Neutrino Astrophysics'' network.

\end{document}